\documentclass[aps,preprint]{revtex4}

\input epsf
\usepackage{amsmath}
\usepackage{amsfonts}
\usepackage{amssymb}%
\usepackage{graphicx}

\newcommand{\bel}[1]{\begin{equation}\label{#1}}
\newcommand{\ee}{\end{equation}}

\newcommand{\beq}{\begin{eqnarray}}
\newcommand{\eeq}{\end{eqnarray}}

\def\be{\begin{equation}}
\def\ee{\end{equation}}

\newcommand{\rem}[1]{}

\newcommand{\cdfmu}{\cite{cdfmu}}
\def\mm{\mu^+\mu^-}
\def\tt{\tau^+\tau^-}
\def\ff{f\bar f'}

\begin{document}

\title{Flesh and Blood, or Merely Ghosts?\\
Some Comments on the Multi-Muon Study at CDF
}
  
\author{Matthew J. Strassler}
\affiliation{Department of Physics and Astronomy,
Rutgers University, Piscataway, NJ 08854}

\begin{abstract}
  A recent CDF paper  suggests (but does not
claim) an anomalous event sample containing muons produced with large impact
parameter, often with high multiplicity and at small angles from one another.
This curious hint of a signal is potentially consistent with the hidden valley
scenario, as well as with some other classes of models. Despite its tenuous
nature, this hint highlights the experimental difficulties raised by such
signals, and merits some consideration. Some of the simplest interpretations of
the data, such as a light neutral particle decaying to muon and/or tau pairs,
are largely disfavored; three-body decays to $\tau^+\tau^-\nu$ appear slightly
better. An alternative speculative possibility --- a ``micro-cascade decay''
--- might be consistent with the data. It is suggested that the
experimentalists involved provide additional plots showing invariant mass
distributions of same- and opposite-sign dimuon pairs, invariant masses of
various classes of displaced vertices, and spatial correlations among vertices
within a cone.
\end{abstract}

\maketitle

\section{Introduction}

A recent CDF paper \cdfmu\ attempts to study a class of unidentified processes, called
``ghost events'', which it fails to explain through known detector or
physical effects.  Among these events \cdfmu\ finds indications or
suggestions (not called ``evidence'') of a new long-lived particle,
with a lifetime of order 20 picoseconds (see Figure 25).  Other plots
in \cdfmu, such as Figures 30 and 36, are intended to suggest that
multiple such particles are produced at the same time within a narrow
cone.

No one would be happier than the author of the present note if this
``suggestion of evidence'' were to hold up under scrutiny.  The hidden
valley scenario \cite{hv1,hv2,hv3}, in which a new hidden sector with
a mass gap is added to the standard model and coupled to it at or
below the TeV scale, naturally predicts high-multiplicity production
of new neutral states, which are potentially very light, and are
typically long-lived, possibly decaying with macroscopically displaced
vertices.  (The scenario has some overlap with the more recent
``unparticle'' scenario \cite{georgi}; in fact unparticle models with
mass gaps are examples of hidden valleys \cite{hvun}.  A recent model
of dark matter \cite{darkhv} is also a hidden valley and has an
interesting variant of the standard hidden valley signatures
\cite{ahw}.)  Indeed, the point of developing the scenario was to highlight
the fact that such models are theoretically reasonable, are allowed by
all known data, and pose significant experimental difficulties.

Unfortunately, although the paper \cdfmu\ is long, it is far too short
given its potential importance, and many critical plots that could
support the case are absent.  Nevertheless, it is interesting and
hopefully useful to consider the implications of its ``suggestion of
evidence''.  Such is the limited goal of this note.  The aim is to
explore a few possible explanations, and to identify analysis
strategies that could be used to clarify the situation, eliminating
certain phenomenological options.  No serious attempt is made to
interpret the data. This exercise may well
be helpful for forthcoming searches at the Tevatron and the LHC, even
if the specific results of \cdfmu\ (and a related attempt at an
interpretation by the experimentalists involved \cite{interpret}) are
eventually discredited.

\section{Preliminary Comments}

We begin with two important 
observations relevant to the discussion below.

First, although the cross-section for ``ghost events'' in \cdfmu\
seems at first glance very large, about 75 pb, the number of new physics
events, if any, is significantly lower, by an unknown amount.  The
number of ghost events, given in Table II of \cdfmu, requires
determination of the efficiency for ``QCD'' effects (that is, QCD
heavy flavor processes), the expected sources of multi-muon events.
This efficiency is inferred in an intricate way from Table I and
Figure 2a of \cdfmu, and its value, $24.4\pm 0.2$ percent, implies
that of the 743006 events passing the dimuon trigger and other
preliminary requirements, 589111$\pm$ 4829 are from ``QCD'' sources.
(This is determined by taking
those 143743 events that pass a tight silicon vertex tracking
requirement and dividing by 24.4 percent.)  Subtracting, one finds
153895 ``ghost'' events.  But if the efficiency estimate
were in error for a subclass of events, and the efficiency were only,
say, 23.4 percent, then the number of ``ghost'' events would drop by
1/5, to only about 128000 events.  Thus the number of ``ghost'' events
is very sensitive to the correctness of the efficiency estimate.

Later in \cdfmu, about 69000 events are attributed to known detector
effects --- and again, by subtraction, one concludes that only 83000
events are unexplained.  Any errors in estimating the detector effects
could also have a drastic effect on the number of events requiring a
new explanation.  For these two reasons, we must view the number of
unexplained ghost events as highly uncertain.

Second, most of the ``ghost'' events, and most of the known
backgrounds quoted in \cdfmu, involve the minimal number of muons (the
two trigger muons), and perhaps one additional muon. Events with at
least two muons in each of two independent cones provide a cleaner
subsample.  If ``ghost'' events with many muons are due to detector
backgrounds, severe failures in tracking or in the muon system might
be imagined as a cause.  But an important correlation is noted in
\cdfmu: for those 27990 events in which one cone of 36.8$^\circ$
($\cos\theta<0.8$) around one of the trigger muons contains an
additional muon, over 10 percent have an additional muon in the cone
around the second trigger muon.  In these 3000 ``2-dimuon-cone''
events, the second cone is generally in a rather different region of
the detector.  To explain this requires an event-wide fluctuation that
can provide correlated fakes in both cones, or a substantial failure
of the modeling of heavy flavor physics, which is claimed in \cdfmu\
to be well under control.  Backgrounds such as decays-in-flight,
punch-through and secondary interactions are not likely to contribute
to both cones.  In short, this subsample is unexpectedly large, and
much cleaner than the full ghost sample.  If there is new
flesh-and-blood physics hiding among the ghosts, this subsample seems
like a good place to look for it.  To explain these events may require
a new process with a cross section closer to 5 -- 10 pb, rather than
100 pb.

\section{High multiplicity of new light neutral particles}

Light neutral particles which can decay to standard model particles,
possibly with long lifetimes, are very poorly constrained by existing
data, as emphasized in \cite{hv1, hv2, hv3, hvwis,hvun}.  Many classes
of models within the hidden valley scenario can generate metastable
particles of spin $\leq 2$, often produced with moderate to high
multiplicity \cite{hv1, hv2,hvun}.  This includes a recent
dark-matter-motivated context \cite{ahw}, where the possibility of
spin-one particles with mass below 2 GeV was particularly highlighted.
Other models can do the same \cite{CFW}.  However, the data pose a
challenge for any attempt to explain the ``ghost'' events using a new
light boson, notwithstanding the interpretation given by a small group
of CDF experimentalists \cite{interpret}.

The data quickly excludes some options.  A new light
particle $X$ which produces muons via the decay $X\to\mm$, or the
decay $X\to\mm$-plus-additional-particles, would generate a strong charge
anti-correlation.
Figure 22 of \cdfmu, which
shows distributions of muon charges within narrow cones, is consistent
with no muon charge correlations or anticorrelations, and strongly
disfavors this interpretation.  A rough estimate and simulated data
both suggest that unless efficiency for muon detection is well below
20 percent or muon fake rates are very large --- both
inconsistent with naive expectations and with estimates given in \cdfmu\
--- new particles decaying directly to $\mm$ would give Figure 22 a
very different appearance.  (Also, a resonant decay
$X\to\mm$ would appear in Figure 34 of \cdfmu; see below.)

The simplest way to get uncorrelated muon charges is to have $X\to\tt$
followed by $\tau\to\mu\nu\bar \nu$.  This then gives an efficiency
for muon {\it production} and detection below 20 percent, and in a way
consistent with the uncorrelated charge distribution give in Figure
22.  This is the approach to the data suggested in \cite{interpret}.
But there are two disquieting --- though not quite fatal --- aspects
of this interpretation, which we will now demonstrate.  The first is
that the distribution of the invariant masses of all muons in a cone,
given in Figure 34, requires that the the ratio
$Br(X\to\mm)/Br(X\to\tt)$ is well below $10^{-3}$, whereas theoretical
expectations would suggest a ratio between 1 (for a typical spin-one
particle) and $m_\mu^2/m_\tau^2=0.0035$ (for a typical spin-zero
particle, such as a scalar mixing with the Higgs boson.)  The second
is that the kinematics of the events shown in Figure 34 are squeezed
into a corner with the minimal amount of available energy, where
$m_X-2m_\tau$ is very small, suppressing the phase space for $X\to\tt$
and making it somewhat surprising that $Br(X\to\tt)$ is large.

To check these statements, various simulations of a Higgs boson
decaying to multiple $X$ bosons were conducted, and a plot analogous
to Figure 34 was generated. Some models were similar to that of
\cite{interpret} though with a different mass spectrum.  Efforts were
made to match the trigger criteria and cuts to the extent possible in
a theoretical simulation.  A plot from one such simulation, with
$m_X=7.2$ GeV, and a branching fraction ratio of
$m_\mu^2/m_\tau^2=0.0035$ for $X\to\mm$ compared to $X\to\tt$, is
shown in Figure 1a.

The figure indicates that the resonance from $X\to\mm$ should have
easily been seen in Figure 34a and 34b, and thus the branching
fraction for this mode must be much less than $10^{-3}$.  This limit
could most likely be improved: Figure 34 uses binning of 200 MeV, but
the dimuon resolution at CDF should be better than this, perhaps by a
factor of 2 or greater.  This is suggested in Figure 1b above.  One
hopes that the experimentalists will update their result to put an
upper bound on $Br(X\to\mm)$ within the context of the hypothesis of a
new light boson.  Also, note that Figure 34 plots the invariant mass
of {\it all} muons in a cone with $\geq 2$ muons. Figure 22 indicates
the plot is dominated by dimuon cones, but only about half of these
have opposite-sign dimuons.  If instead the plot were limited to
opposite-sign dimuons, the continuum background would drop and the
limits might further improve.

\begin{figure}[htbp]  \begin{center}  \leavevmode
\vskip 0.5in
\includegraphics[width=.9
\textwidth]{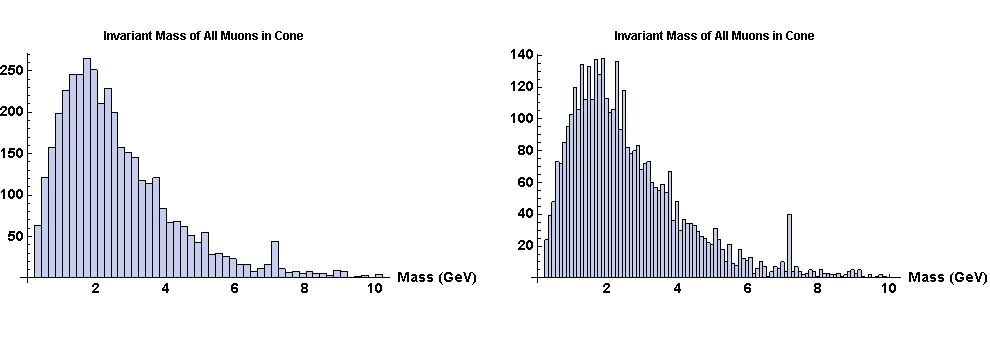} 
\end{center}
\vskip -0.00in \caption{ Analogous to Figure 34a of \cdfmu, the
invariant mass of all muons within cones that contain two or more
muons for the toy model described in the text, with a 7.2 GeV
particle $X$.  (a) 200 MeV bins, as
in Figure 34a of \cdfmu; (b)
100 MeV bins, highlighting the $\mm$ resonance at $m_X=$ 7.2 GeV.}
\label{H16t6}
\end{figure}

In figure 1a one also sees a substantial tail, up to and beyond the
$X$ mass, from the dominant $X\to\tt$ decays.  The absence of such a
tail in Figure 34 of \cdfmu\ implies the $X$ mass cannot be much more
than 3 GeV.  Of course, it must be heavier than $2m_\tau=3.55$ GeV.
Thus the kinematics from Figure 34 forces $m_X$ toward its lowest
possible value.  (While the low-mass end of Figure 1 also fails to
match Figure 34 of \cdfmu, one should probably disregard this fact,
since this region can more easily be affected by mismatches between
the crude theoretical study and the details of the CDF detector.)

Indeed, to fit the data, the interpretation \cite{interpret} actually
requires $m_X\approx 3.6$ GeV.  This reduces the phase space for
$X\to\tt$ by a factor of 30 or more compared to naive estimates, and
thus would increase the branching ratio for $X\to\mm$ (or anything
else) by a corresponding factor.  To get a sufficiently small
branching fraction for $X\to\mm$ then implies that the $X$ coupling to
muons must be smaller than expected for a typical scalar by at least a
factor of 10.  Similarly, any decay to hadrons, such as $X\to K^+K^-$,
would also be relatively enhanced due to the small $X\to\tt$ phase
space, so $X$ couplings to quarks must be very small.

Thus an $X\to\tt$ interpretation of \cdfmu\ forces on us a new particle that
couples to $\tt$ only.  Since tau-number is violated by the mixing of tau
and muon neutrinos, adding such a particle
risks introducing unobserved flavor-changing-neutral-current tau decays.

Here are three additional comments on the specific interpretation suggested in
\cite{interpret}.
\begin{itemize}
\item The long lifetime of the $X$ (called $h_3$ in \cite{interpret})
could arise in a number of ways.  But it cannot arise from the small
splitting between $m_X$ and $2m_\tau$.  In the limit $m_X\to 2m_\tau$,
the tau-loop-induced decay $X\to \gamma\gamma$, which cannot be
forbidden, will always come to dominate, eliminating the muon
signal.  
\item 
To avoid the unseen high-side tail in Figure 34, \cite{interpret} is
forced to take not only the mass of $h_3$ to be close to $2m_\tau$, but
also $m_{h_2}\approx 2m_{h_3}$ and $m_{h_1}\approx
2m_{h_2}$.  In short, to fit Figure 34 requres a triple fine-tuning
into the kinematic corner with the minimal visible energy.  While possible,
it hardly seems likely.
\item The fact that the model lies in this kinematic corner should
make it relatively easy to verify or falsify with the existing data.
In the cascade decay $h_1\to 2h_2\to 4h_3$ suggested in
\cite{interpret} to explain the multi-muon cones in \cdfmu, the small
energy available in the cascade decay implies the relative velocity of
the four $h_3$ particles in their shared center-of-mass frame will be
small.  However, their velocities in the lab frame must be large to
generate a 3 GeV trigger muon.  {\it Therefore, the four $h_3$
particles will be tightly collimated, which predicts that all four
displaced vertices from the $h_3$ decays will lie approximately on a
line pointing back to the collision point.}
\end{itemize}

If the data does not support four highly collinear decay vertices, or
we simply find the fine-tuning of \cite{interpret} akin to epicycles,
what else might be responsible for the data?  There are certainly
other options, the simplest being a fermion $X$ instead of a
boson, so that it might decay to $\tt\nu$ and $\mm\nu$.  (For example,
this could happen in non-minimal supersymmetry with R-parity
violation.) This eliminates the resonance feature of Figure 1 (turning
it into an unmeasurable enhancement of the continuum) and pushes the
continuum of Figure 1 down to lower values.  In this case, theoretical
studies suggest that $m_X$ could perhaps be raised to as much as 4.5
or 5 GeV, so the kinematic fine-tuning required for the
models of \cite{interpret} to fit the data is certainly not quite as severe.
On the other hand, a problem pointed out by the authors of
\cite{interpret} is the inability of their simplest models to obtain
the large track-$p_T$ seen in Figure 32 of \cdfmu.  If neutrinos carry
off even more of the energy, it potentially makes this worse.  

One should also consider the skeptic's interpretation of Figures 22
and 34.  The uncorrelated muon charges are consistent with random
hadronic tracks being misidentified as muons, random tracking errors,
etc. Also, Figure 34 is potentially consistent with a random
distribution shaped mainly by triggering and cuts.  A trigger muon
(with $p_T\geq$ 3 GeV) combined at random with a non-trigger muon
(with $p_T\geq$ 2 GeV) within a cone of $\cos\theta<0.8$ will have an
invariant mass of order or below 1.5 GeV, potentially consistent with
the distribution in Figure 34.  This observation might motivate
raising the $p_T$ cuts to see whether the low-invariant-mass region
entirely disappears.

There would seem to be a natural check of the whole story (beyond the
specific interpretation of \cite{interpret}.)  If there were a physics
signal from new light particles, then the dimuon-invariant-mass
distribution in dimuon cones would show two components: (I) from
opposite-sign dimuons from the same $X$ decay, and (II) from opposite-
or same-sign-dimuons from different $X$ decays.  By restricting the
plots in Figure 34 to dimuon cones, and separating them into same- and
opposite-sign dimuons, one could measure (I) and (II), determine
whether in fact (I) is present, and estimate the $X$ mass directly
from the data.  (If in fact there were no signal one would expect no
evidence for the component (I); opposite- and same-sign distributions
would look the same.)  Armed with this information, one could then
perform a critical cross-check. Opposite-sign muon pairs with low
masses (below $m_X$) would be much more likely to stem from a two-muon
displaced vertex than either same-sign pairs of any mass or
opposite-sign pairs with higher invariant mass.  It is not clear why
these checks were not performed in \cdfmu.

There are a number of other plots whose presence, or absence, in
Appendix B of \cdfmu\ is very surprising.  In particular, though
obviously presented so as to support the interpretation of
\cite{interpret}, the plots of Appendix B do not actually appear to do
so.

As far as can be discerned from the text of \cdfmu\ and the captions of
Figure 45 and 46, it appears that locations and invariant masses of
three-track combinations are presented in order to support the idea
that the events are rich in taus.  But this seems odd: if the
interpretation given in \cite{interpret} is correct, any three-prong
tau is emitted in an $h_3$ decay that has at least one more track.
Therefore, any three-track vertex (if real) is actually a four-track
vertex where one track has been lost or otherwise not included.  At
best, the combinatoric background for the search is therefore 3 to 1.
At worst, since tracks in three-prong tau decays are softer than in
one-prong decays, the probability that the three tracks observed
actually come from a single tau is much less than 1/4.  Finally, there
is the issue of cuts sculpting the distribution.  If one selects three
tracks with $p_T>1$ GeV from within a cone of $\cos\theta<0.8$, the
invariant mass of those tracks will typically be of order 1.1 GeV,
close to the expectation for 3 tracks from a three-prong decay, and at
the peak of the observed distribution. The evidence that this
sculpting effect is unimportant seems insufficient.  Clarifying
comments from the authors of \cite{interpret} would be welcome.

Conversely, the interpretation of \cite{interpret} would imply that
opposite-sign dimuon vertices should be quite common (appearing in 3
percent of vertices and 12 percent of cones), should have relatively
low backgrounds, and should have a very distinctive invariant mass
distribution.  The location of such vertices is shown in Figure 44 of
\cdfmu, but inexplicably the invariant mass distribution is not given.
Clearly these plots could have
potentially supported the case that a signal is present in \cdfmu, and
perhaps even the interpretion of \cite{interpret}.
Similar plots for same-sign dimuon vertices and for muon-plus-track
vertices would also be instructive.

For the study in Appendix B, the $p_T$ cuts are taken very low,
presumably to obtain high statistics.  But would it not have been
better to focus on higher-$p_T$ tracks in events with 2-dimuon-cones,
obtaining a much cleaner sample with much lower detector backgrounds?
The advantage of interpretability would seem likely to outweigh the
cost in statistics.

More generally (as we will see below) there is every reason to
consider vertices with $k$ tracks, $k$ any integer.  To choose a
particular $k$ in advance is to risk biasing the study toward a
particular interpretation, rather than allowing the data to speak for
itself.  It would be helpful to have a much more systematic, and less
targeted, study of the vertices in these events.

A final issue with this analysis, not entirely robust but still worth
mentioning, arises if the $X$ particle can sometimes be produced at
low multiplicity $n_X$.  (This issue would not arise in the
interpretation \cite{interpret}.) In a low $X$-multiplicity cone,
there is a certain probability that only one $X$ will have significant
$p_T$; the others (if any) may be semirelativistic, with $p_T$ or
order $m_X$.  In this case the decay products of the slow $X$ bosons
may be so soft or at such large angles that they will not affect the
isolation criteria of the daughters of the hard $X$.  A search for a
single isolated well-displaced $\mm$ pair, with 380 pb$^{-1}$ of data,
was carried out by DZero in \cite{TAdams}.  No candidate events were
observed.  This should strongly constrain the number of $n_X=1$ cones,
and probably even $n_X=2$ cones, within the ``ghost'' sample.


In summary, the data seem to exclude high-multiplicity production of
several types of light particles.
\begin{itemize}
\item Particles decaying often to $\mm$ are very strongly disfavored, by Figures 22 and 34.
\item Particles decaying often to $\mm\nu$ are strongly disfavored, by Figure 22.
\item Particles decaying to $\tt$ and to $\mm$ are strongly
disfavored for $Br(X\to\mm)> 0.001$, by Figure 34.
\item Particles decaying to $\tt$ and not to $\mm$
are disfavored by Figure 34 unless
$m_X$ is uncomfortably close to $2m_\tau$.
\item Particles decaying to $\tt\nu$, and possibly $\mm\nu$ with a
small branching fraction, may be allowed by the data if $m_X$ is not too 
far above $2m_\tau$.
\end{itemize}
In any case, it remains to explain the $p_T$ distribution implied
by Figure 32 of \cdfmu.

Of course more complicated decays (such as direct four-body
decays, or an intricate and non-minimal spectrum of $X$ particles)
might also be allowed by the data.  But to proceed any further, more
experimental information is needed.

\section{Micro-cascades}

While awaiting the improved experimental constraints on the $X\to \tt(\nu)$
option, it is interesting to note a rather different approach to
understanding the data: a ``micro-cascade''.  Consider a set of
relatively heavy particles $P_i$, $i=1,\dots,n$, with small mass
splittings; take $P_n$ to be the heaviest and $P_1$ the lightest.
Imagine these decay one to the next by $P_i\to f\bar f' P_{i-1}$,
emitting a pair of standard model quarks or leptons.  Experimentally,
each decay would generate a pair of soft dileptons, or a soft lepton
and a neutrino, or perhaps soft hadrons.  This sequence of decays or
``micro-cascade'' is sketched in Figure 2.  There are no theoretical
obstructions to such a phenomenon, nor any existing data that could
exclude it.  Indeed it is possible to construct extensions of the
standard model, and especially easy to build hidden valley models,
that could give this signature.  

While many details will vary from one model to the next, the most
reliable prediction of such a phenomenon stems from the fact that
the heavy particles do not suffer
much of a change in their velocity as they decay one to the next.
This means that {\it the vertices from the decays are aligned,
roughly, in a straight line} (though possibly bent slightly by the
magnetic field if the heavy particles are charged.)  If multiple
tracks are emitted at each vertex, this should be relatively easy to
verify. If only one track is emitted (or, on average, reconstructed)
at each vertex, then a novel track correlation study is needed to
check whether the various displaced tracks in a cone all intersect a
single line pointing back to the primary vertex.

However, as noted above, this same feature (multiple
collinearly-aligned vertices) is also a property of the kinematically
squeezed $h_1\to2h_2\to4h_3$ decays suggested in \cite{interpret}.
To distinguish the two requires a more careful study
of the correlations of the {\it radial} location of the vertices.
{\it Thus both
angular and radial correlations among vertices are key observables in
any interpretation involving multiple nearby displaced vertices,} and
it would be useful for any analysis aimed at such a signal to consider
these observables.

\begin{figure}[htbp]  \begin{center}  \leavevmode
\vskip 0.5in
\includegraphics[width=.75
\textwidth]{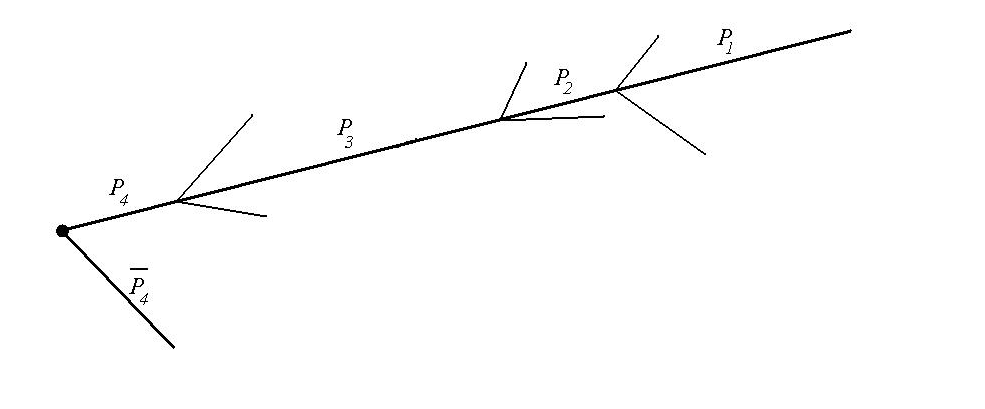} 
\end{center}
\vskip -0.00in \caption{ Schematic view of a micro-cascade.  Each
heavy particle $P_i$ decays to a particle $P_{i-1}$, with a relatively
small mass splitting, while spitting off standard model leptons or
hadrons.  A sequence of such decays can create a line of displaced
vertices.  }
\label{microcas}
\end{figure}
    
\subsection{The various types of micro-cascade}

There are three questions whose answers determine much of a micro-cascade's
phenomenology.  (1) What are the standard-model charges of the heavy
particles $P_i$? (2) What are the branching
fractions for the decays $P_i\to \ff P_{i-1}$ for various choices of
$f\bar f'$?  (We assume for simplicity that each $P_i$ decays predominantly
to $P_{i-1}$ and not to lighter $P_j$.) (3) What is the fate of $P_1$?  Let us address these in 
turn.

If the $P_n$ carry color, then (since the
displaced vertices require they are long-lived) they will hadronize,
generating occasional stray hadronic tracks (from fragmentation and
hadronic decays) both at the primary vertex and at subsequent vertices.
Otherwise, they will behave like heavy leptons or neutrinos.
Meanwhile, if they form nontrivial electroweak $SU(2)$
multiplets, they can potentially decay, just as do
standard model quarks and leptons, by $W$ boson emission.  The decays
occur at tree-level and, given the mass splittings and CKM-like
mixing angles, are calculable.  Branching fractions are 
determined by couplings to the $W$ and by kinematics.
If instead the $P_i$ are $SU(2)$
singlets, then they must decay, perhaps through loop effects, by
emission of neutral particles, such as 
off-shell $Z$ bosons or Higgs bosons, or a new unknown particle.
(If photon emission is
allowed, then it typically dominates,
as in the case of $b\to s$ decays; given
the signature suggested in \cdfmu, let us assume photon decays are absent.)
Decay rates tend to be highly model-dependent.
So do branching fractions, though they are most likely to be
either generationally-democractic (as for a $Z$ or typical $Z'$)
or weighted by mass-squared (as for a Higgs or new scalar/pseudoscalar.)

Finally, there are four natural
options for the fate of the $P_1$ at the end of the cascade:
\begin{itemize}
\item $P_1$ is neutral, weakly interacting, and stable on detector
time-scales; it exits the detector unseen.
\item $P_1$ is charged, electromagnetically interacting, and stable
on detector time scales; at first glance it appears to be a muon.
\item $P_1$ is colored and forms hadrons with light quarks; the stable
or metastable hadrons formed may be neutral, escaping unnoticed until
searched for, or may be charged, often masquerading as a muon, or
both.
\item $P_1$ is unstable and decays within the detector volume, providing
one last vertex in the cascade, possibly with many tracks and/or 
considerable visible energy.
\end{itemize}
A stable $P_1$ will exit the detector typically within the cone around
the nearest trigger muon.  A stable neutral $P_1$ leaves some amount
of missing transverse momentum, which unfortunately tends to cancel
between the $P_1$ and $\bar P_1$ in an event.  A stable charged
particle will actually add to the muon count within the cone, and
might show up through precise time-of-flight measurements.  New
hadrons, both neutral and charged (and even charge-flipping), might be
observable through their unusual interactions with matter.  Finally, a
an unstable $P_1$ would provide the terminal vertex in a
micro-cascade. Much or all of the energy from its decay would be
recorded, in contrast to the other cases where the stable $P_1$ would
leave little or no energy in the calorimeter.

All of the above options are interesting in that they could give
challenging experimental signatures, and only some of them (mainly
stable charged particles) are on the usual experimental analysis menu.
In the case at hand, a first glance at \cdfmu\ turns up nothing which
directly suggests or disfavors any one of these phenomenological
possibilities.  

Regarding the production of the $P_i$, there are several options,
given the fairly large cross-sections needed to explain \cdfmu.  For
colored $P_i$ particles, ordinary pair production may be sufficient.
However, pair production predominantly creates $P_i$ moving with
rather low velocity, in which case the leptons and hadrons produced in
the $P_i$ decays would go off in all directions.  To collimate these
particles into narrow cones, as seems required to explain \cdfmu,
requires a boost of the parent $P_n$.  For this reason, the data
motivates consideration of models where the $P_i$ appear in the decay
of a heavy resonance $R$.  In this case the $P_i$ production rates are
determined by their couplings to the resonance, allowing large
cross-sections even for color-neutral $P_i$, and also the $P_i$ are
somewhat boosted, if
they are light compared to $m_R/2$.
Still, pair production of colored $P_i$ is by no means excluded by the
discussion below.

\subsection{Brief aside on model-buildinng}

It should be stressed that there is nothing exceptionally natural
about a micro-cascade.  It could easily be the case that all $P_i$
directly decay to the lightest particle $P_1$.  It might be that
lifetimes are different by several orders of magnitude, implying that
only one or two displaced vertices, with a high multiplicity of
tracks, can actually be resolved.  For this reason, detailed
model-building seems premature; let us see if the data actually shows
a signature.  Still, for the interested reader, a few preliminary
comments are in order.

Despite some risk of ruining perturbative unification of the
standard model gauge couplings, it is straightforward to add several
new Dirac fermions or scalars at the weak scale, charged in one way or
other under the standard model gauge couplings, without violating
experimental constraints.  More precisely, this is true as long as
these particles do not get their masses mainly from the Higgs boson
(which would affect electroweak precision measurements) and are heavy
enough to have avoided direct discovery at LEP.  (Indeed this is what
could happen in certain supersymmetric or extra-dimensional models if
their mass spectrum were squeezed.)  It is even simpler to add
standard-model-neutral particles (as in a hidden valley); such
particles could have any mass.

Near-degeneracies and long lifetimes can arise even in minimal
supersymmetric models, as in the limit that $M_2\ll M_1$ (e.g. anomaly
mediation) or $\mu\ll M_2,M_1$ (light Higgsinos).  In a non-minimal
model of supersymmetry, one may obtain a richer near-degenerate
spectrum.  For instance, if gauginos are Dirac instead of Majorana,
the spectrum is doubled, and small splittings are induced if the
Majorana terms are small.  Similar degeneracies could easily arise in
extra-dimensional models, depending on the shape of the extra
dimensions.

One easy way to obtain a near-degenerate spectrum with a micro-cascade
is to obtain a large multiplet of a weakly-broken symmetry.  For
example, consider that in QCD, if the $b$ quark had a lifetime of
seconds, then the $B_u,B_d,B_s,B^*_u,B^*_d,B^*_s$ system would form six
near-degenerate states that could only decay through electroweak
interactions.  Were there no photon, all of their decays would all be
through soft lepton emission.  A new heavy vectorlike long-lived quark of QCD,
added to the standard model, would have such
a spectrum, but it would be too degenerate for present needs.  However, a
confining hidden valley with a similar structure 
of v-quark masses would have near-degenerate v-mesons that could only
decay to standard model particles and that might be able 
to explain \cdfmu.  (If the heavy v-quark were
in an antisymmetric tensor representation of hidden-color rather than in a
fundamental, then the number of near-degenerate states would
double.)  
 A scalar that has hidden-flavor-changing couplings and mixes
with the Higgs boson could then allow these hidden states to decay to
one another.  If the splittings all lie between $2m_\tau$ and $2m_b$,
the dominant final state would be non-resonant $\tt$ pairs.

In this example, the heavy v-quark inside the v-meson, and therefore the
v-meson itself, could potentially carry standard model color.
This would realize the option  of colored $P_i$ particles, and allow
naturally large cross-sections without a new resonance.

Another way to obtain a micro-cascade would involve a weakly-coupled
hidden valley with an extra dimension of radius $R\equiv1/\mu$.  A
massive five dimensional particle with a large mass $M$ would then
have a tower of states $P_i$ with splittings of order $\mu^2/M$.  (A
similar model could be built using the spectrum of a string with
massive ends.)  If a scalar $S$ violates all conserved Kaluza-Klein
charges, then particles in this tower can decay via off-shell $S$
emission.  Mixing of $S$ with the Higgs boson would then allow $\tt$
final states in a cascade.

One could and eventually should do a much better job of model-building than
attempted here.  Let us now return to the data and phenomenology.

\subsection{Reducing the space of options: I}

What if most of the displaced muons came from three-body $P_i\to \mm
P_{i-1}$ decays?  
As before, the lack of muon charge correlations in
Figure 22 would be reasonable only if the
efficiency for muon detection were implausibly low.  We therefore set
this option aside.

Consistency with Figure 22 would be much improved if the $P_i$ decay via
off-shell $W$ bosons.  A decay $P_i\to \mu^+\nu P_{i-1}$ is just as
likely to be followed by $P_{i-1}\to e^-\bar \nu P_{i-2}$ or
$P_{i-1}\to \rho^- P_{i-2}$ as $P_{i-1}\to \mu^-\nu P_{i-1}$; there is
no correlation between the final states of the first and second decay.

Note that even if $P_i\to P_{i-1}\ell\nu$ is the dominant decay mode,
so that many steps in the micro-cascade produce only one track, it may
still be the case that at least two tracks emerge at each displaced
vertex.  (The efficiency to detect these tracks is a separate issue.)
This is because in this scenario there is some conflict between making
the splittings between the $P_i$ large enough to generate triggerable
muons in the decays while still permitting all the $P_i$ lifetimes to
be of order a few ps or more.  Instead there is a tendency, as in the
decays of standard model quarks, for long lifetimes to require small
off-diagonal mixings, in which case the diagonal decays are rapid.
For example, a slow $P_3\to \mu^+\nu P_2$ decay might be likely to be
followed by a faster $P_2\to e^-\bar \nu P_1$ decay.  Such a chain
might be detected as a $\mu^+e^-$ pair appearing at a single point.
Note that $\mu^+e^+$ vertices would not occur.

Thus, as a result of these overlapping decay steps in the micro-cascade,
many of the muons may be produced at a {\it composite decay vertex},
with one or more additional tracks.  (Note that if the $P_i$ are
colored and form hadrons, sometimes unstable, in each decay, the
number of tracks at each step may be further enhanced.)  Consequently
it would be helpful if an update of \cdfmu\ would give more
information about the vertices, including a histogram of the number of
tracks that are present in vertices with 0, 1, or 2 muons, and how the
number of vertices within a cone is correlated with the number of
muons in that cone.

This phenomenological scenario may not work for \cdfmu, however.  The
particles $P_i$ must be fairly light, if they are to be produced
abundantly and with a reasonable boost.  But if they are $SU(2)$
non-singlets, some of them must be electrically charged, which
presumably requires their masses to be close to or above 100 GeV (as
some searches for displaced decays were carried out at LEP II).
Pair production of heavy colorless particles has a low
cross-section, while obtaining high enough rates from a new resonance
well above 200 GeV might not be possible.  This requires additional
study.

\subsection{Reducing the space of options: II}

If instead the $P_i$ are colored and are all electrically neutral, the
constraints on their masses are much weaker, since they would not be
produced at LEP; masses of 50 GeV \cite{becherschwartz} are apparently
allowed.  (However such particles generally ruin coupling-constant
unification.)  If the $P_i$ are neutral under the entire standard model,
as in hidden valleys,
then there are hardly any constraints.

As mentioned above, examples of hidden valleys with heavy neutral or
colored particles $P_i$ decaying via a three-body decay to $f\bar f
P_{i-1}$ can be obtained through multi-flavor generalizations of
examples discussed in \cite{hv1}, and many other approaches.  (Events
from a similar model, with larger mass splitings than relevant here,
were shown in \cite{fourflavor}.)  Other types of models can
also generate these phenomena.

Decays in this case must be of the form $P_i\to \ff P_j$, where $f$
and $f'$ have the same charge.  We saw already that Figure 22
disfavors $P_i\to \mm P_{i-1}$ as an important source of muons, but it
still allows $P_i\to \tt P_{i-1}$ to be the dominant souce.

\subsection{A toy example}

Consider a toy example, which is intended to illustrate some simple
observations (rather than as a serious attempt to fit the data.)  Here
a 200 GeV resonance decays to $P_5\bar P_5$ is considered, where $P_5$
has mass 60 GeV; this then decays in a cascade $P_5\to P_4\to P_3\to
P_2\to P_1$, where $m_4 = 53$ GeV, $m_3 = 45$ GeV, $m_4 = 38$ GeV,
$m_1 = 31$ GeV.  The $P_i$ are assumed all to be standard-model
neutral.  All decays are assumed to occur via $P_i\to \tt P_{i-1}$.
The plot corresponding to Figure 34a in \cdfmu\ is shown in Figure 3a.
Because the decay is three-body, the result is closer to satisfactory
than in Figure 1, and indeed is similar to the case $X\to\tt\nu$
described in the previous section.  The plot suggests that a mass
splitting among the $P_i$ of order 7 -- 8 GeV is a bit too large,
instead favoring a mass splitting closer to 4 -- 5 GeV.

Given this moderate success, what fails?  Again the $p_T$ distribution of the
tracks is too low to match Figure 32 of \cdfmu.   This is
no surprise, since a 200 GeV resonance will not often give cones with
tracks that have $\sum p_T> 60$ GeV, {\it especially} if the $P_1$ is
invisible and stable.  But if instead the $P_1$ decays in flight, its
decay products could both increase the observed $p_T$ and the observed
number of tracks.  However, the effect depends crucially on how many
tracks $P_1$ decays to, and also on the $P_1$ lifetime, since
sufficiently displaced or angled tracks might not all be
reconstructed.  To study this is beyond a theorist's capability; a detailed
understanding of the detector is needed.



For figure 3a, it was assumed that $P_i\to \mm P_{i-1}$ is disallowed.
The effect of including this decay mode with a branching fraction
of order $m_\mu^2/m_\tau^2$ is shown in Figure 3b.  Comparing with
Figure 3a we see little change.  Thus, as in the case $X\to\tt\nu$,
Figure 34 of \cdfmu\ can be matched without
making the branching fraction to muons unnaturally small.


The toy model thus indicates that a micro-cascade is not obviously any
worse at explaining the data than is high-multiplicity production of a
light particle.  In general, micro-cascades should be considered as an
option whenever multiple long-lived particles are suspected.

\begin{figure}[htbp]  \begin{center}  \leavevmode
\vskip 0.5in
\includegraphics[width=.9
\textwidth]{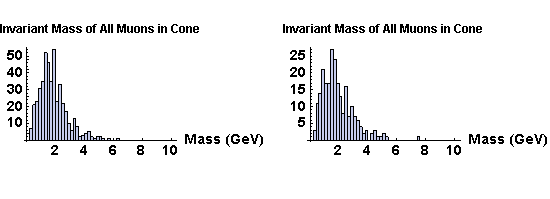} 
\end{center}
\vskip -0.00in \caption{ For the micro-cascade toy model, with
mass splittings of order 7--8 GeV, a plot
analogous to Figure 34a of \cdfmu.  (a) Decays via $P_i\to\tt P_{i-1}$
only.  (b) Decays by $P_i\to\tt P_{i-1}$ and (with low branching
fraction) $P_i\to\mm P_{i-1}$.  (The lower statistics is a consequence
only of the simulation and has no physical significance.)}
\label{R34}
\end{figure}

\subsection{Summary on micro-cascades}
A summary of the signatures for a micro-cascade are the following:
\begin{itemize}
\item
the model-independent signature of a chain of vertices, one after
another (or of a common line that points back to the primary
vertex and is crossed by all the displaced tracks);
\item correlations in the radial positions of vertices due to the
varying lifetimes of the new particles $P_i$ that participate in the cascade
\item possibly unexpected and widely varying combinations of particles
emerging from the vertices (due to spatial overlap of multiple decay
steps, and/or to hadronization effects from colored $P_i$ decays)
\item possible ``terminating vertices'' at the end of the micro-cascade, where
the last of the cascading particles decays with more energy and perhaps
tracks than at any of the vertices within the micro-cascade;
\item possible long-lived charged particles, weakly or strongly interacting,
masquerading as muons and pointing back to the micro-cascade vertices;
\item possible long-lived exotic hadrons, charged or neutral, pointing
back to the micro-cascade vertices.
\end{itemize}
Of course, there are existing Tevatron studies looking for long-lived
charged particles \cite{CDFch,DZeroch}.  
These, along with the combination of mass
constraints from LEP and the need for large cross-sections, may
already have excluded the last two possibilities in
the regime of interest.  The other options lie somewhat outside the
usual realm of particle physics analyses.  It will be interesting
to see whether they can be carried out successfully in any update
of \cdfmu.

\section{Final comments}

While considerable skepticism is still in order, there is still some
hope that \cdfmu\ contains a hint of a new phenomenon.
The considerations of this note suggest some additional plots that
might be useful to include in any update of \cdfmu.
\begin{itemize}
\item A breakdown of Figures 34a and 34b into subcases: opposite-sign
dimuon, same-sign-dimuon, and multi-muon.
\item Figure 34 (and its subcases)
with narrower binning, appropriate to the CDF mass resolution.
\item Plots showing the correlation, within a single cone around a
trigger muon, between the numbers of muons, numbers of non-muon
tracks, and numbers of vertices.
\item Plots showing the number of tracks and 
the invariant mass of all tracks within a cone,
for cones with different numbers of muons.
\item Plots similar to Figures 45 and 46 for a wider variety of vertex types,
and plots comparing same-sign and opposite-sign di-track (or di-muon or muon-plus-track) vertices.
\item Plots showing the number of tracks per vertex for vertices
containing 0, 1 or 2 muons.
\item Plots studying  the radial and angular
correlation between multiple
vertices in a cone.
\end{itemize}
In each of these cases, it would be important to see these plots
for the cleaner subsample of 3000 ``ghost'' 2-dimuon-cone events.

Separately, considerable light might be shed using
those events where some muons pass
the loose SVX criterion.  For
example, for the 3000 ``ghost'' 2-dimuon-cone events, how many have at
least one pair of muons that pass the loose SVX criterion and form a
dimuon vertex?  how many have at least two muons passing the loose SVX
criterion that each have a muon-plus-track vertex?  A substantial number
should be expected, if there really is a new particle of lifetime
$\sim$ 20 ps.  According to the text of \cdfmu, the loose SVX criterion
``accepts muons from parent particles with a decay length as long as
10 cm''.  Since $c\tau\sim 0.6$ cm for $\tau = 20$ ps, even a
substantial boost of order 10 for the decaying particles will still
leave many decays at considerably shorter distances.  If we crudely
estimate that (a) 1/2 of the cones in Figure 34b have opposite-sign
dimuons, and (b) 1/4 of these dimuons come from the same decay step,
then accounting for both cones suggests hundreds of events have
vertices built from loose-SVX muons.  A careful study of this cleaner
subsample, on its own merits, and also comparing it with
the sample of events with no SVX requirement, would be most
illuminating.

Then there is the issue of the electrons.  While much more difficult
to observe than non-isolated muons, and while suffering from large
conversion backgrounds, the non-isolated electrons that almost any
reasonable model of the new physics would predict must be identified.
At least, an analysis of the electrons and positrons found in a small
low-background sample of multi-muon events, such as the 2-dimuon-cone
event sample, would be valuable.  How many electrons there are, how
their charges correlate or anticorrelate with the muon charges, how
many vertices they participate in, {\it etc.}, are crucial questions
requiring experimental answers.  Comparison of same-sign and
opposite-sign $\mu e$ vertices --- the latter being predicted by
$X\to\tt(\nu)$, most  micro-cascades, and
other physics signals, with a sign-uncorrelated background from secondary
interactions of muons --- would be especially valuable.

Finally, even if the hints in \cdfmu\ turns out to be as ephemeral as a
ghost, the challenges that this analysis faces are useful as a
springboard for discussion.  Clearly, if there were a signal of this
type in the data, it would indeed be quite difficult to find it, and
the approach used in \cdfmu\ is far from optimal.
Opening our minds regarding the possible signatures that nature might
provide, and finding new techniques for expanding the range of reasonable
searches at hadron colliders, is surely beneficial for the field.

\

-----------------------

\

I am grateful to my colleagues E. Halkiadakis and A. Lath for helping
me to understand some details of the CDF detector and some of the
methodology used in \cdfmu, and to Y. Gershtein and
S. Thomas for useful discussions.

\end{document}